\documentclass[runningheads]{llncs}
\usepackage{graphicx}
\usepackage{amsmath,amssymb}
\begin{document}
\title{Transitions among metastable states underlie context-dependent working memories in a multiple timescale network}
%
%
\author{Tomoki Kurikawa\orcidID{0000-0003-1475-2812}}
\authorrunning{T. Kurikawa}
%
\institute{Department of Physics, Kansai Medical University, Shinmachi 2-5-1, Hirakata, Osaka, Japan
\email{kurikawt@hirakata.kmu.ac.jp}}
\maketitle              
\begin{abstract}
Transitions between metastable states are commonly observed in the neural system and underlie various cognitive functions such as working memory. In a previous study, we have developed a neural network model with the slow and fast populations, wherein simple Hebb-type learning enables stable and complex (e.g., non-Markov) transitions between neural states. This model is distinct from a network with asymmetric Hebbian connectivity and a network trained with supervised machine learning methods: the former generates simple Markov sequences. The latter generates complex but vulnerable sequences against perturbation and its learning methods are biologically implausible. By using our model, we propose and demonstrate a novel mechanism underlying stable working memories: sequentially stabilizing and destabilizing task-related states in the fast neural dynamics. The slow dynamics maintain a history of the applied inputs, e.g., context signals, and enable the task-related states to be stabilized in a context-dependent manner. We found that only a single (or a few) state(s) is stabilized in each epoch (i.e., a period in the presence of the context signal and a delayed period) in a working memory task, resulting in a robust performance against noise and change in a task protocol. These results suggest a simple mechanism underlying complex and stable processing in neural systems.

\end{abstract}
\section{Introduction}
Neural trajectories are commonly observed in neural systems \cite{Miller2016} and are involved in temporal information processing, such as working memory\cite{Stokes2013} decision-making \cite{Kurikawa2018}.
The organization and processing of task-related information through these trajectories is an essential question in neuroscience.

Two theoretical approaches are proposed to answer the question.
In this approach, each pattern in the trajectory is represented as a metastable state, wherein synaptic connectivity in a neural network is formed through Hebbian learning.
An asymmetric connection from the current to the successive pattern\cite{Amari1972,Kleinfeld1986,Sompolinsky1986,Seliger2003,Recanatesi2015,Russo2012,Gros2007} leads to transitions between patterns.
These sequences of metasable states are robust to noise and are widely observed in neural systems\cite{Miller2016}.
A transition between these states is explicitly embedded into connectivity (i.e., the connectivity composed of correlations between the current and the next patterns), 
resulting in a next pattern being determined only by the current pattern.
Hence, the generation of non-Markov sequences depending on the long history of previous patterns is not possible.

In another approach\cite{Mante2013,Chaisangmongkon2017,Sussillo2009,Laje2013} trained recurrent neural networks (RNNs) were used to generate neural trajectories using machine learning methods.
RNNs reproduce neural trajectories observed in neural systems, and how these trajectories encode and process information over time has been investigated.
These models allow for generating complex sequences dependent on the history.
However, parameters are finely tuned through non-biologically plausible learning, and the parsimonious principle of temporal information processing in neural dynamics is unclear.
Thus, a simple and biologically plausible model that generates long-history-dependent sequences is necessary.

To this end, we have studied a neural network model with fast and slow neural dynamics\cite{Kurikawa2020a},  wherein the slow dynamics integrate previous information and regulate fast dynamics.
Using this model, in this study, we propose a novel mechanism underlying stable working memories: stabilizing and destabilizing task-related states at an adequate time.
We also examine whether our model performs a context-dependent working memory task wherein history-dependent computation is necessary.

\section{Model}
\subsection{Neural dynamics}
Our model is based on a previous study\cite{Kurikawa2020a} that showed that an RNN with multiple timescales enables the learning of sequential neural patterns including non-Markov sequences.
The model has two populations with different timescales connected to each other (Fig. 1A). 
One population comprises $N$ fast neurons $x$, and the other comprises $N$ slow neurons $y$.
The fast population receives an external stimulus and generates a target response corresponding to the given input.
These neurons evolve according to
\begin{align}
    \tau_{x}\dot{x_{i}} &= \tanh{(\beta_x I_{i})} - x_{i},  \\ \label{eq:neuro-dyn}
    \tau_{y}\dot{y_{i}} &= \tanh(\beta_y x_{i}) - y_{i},  \\ \label{eq:neuro-dyn-slow}
  I_i  &= u_i  +  \tanh(r_i) +(\eta^{\alpha}_{\mu})_{i},\\
  u_{i} &= \sum_{j \neq i}^{N} J_{ij}^{X} x_{j},\\ 
  r_{i} &= \sum_{j \neq i}^{N} J_{ij}^{XY} \tanh(y_{j}),
\end{align}
where $J_{ij}^{X}$ is a recurrent connection from the $i$ to $j$-th neuron in the population of $x$ and $\boldsymbol{J}^{XY}$ is a connection from the $i$-th neuron in the population of $y$ to the $j$-th neuron in the population of $x$.
The mean values of $J^{X}$ and $J^{XY}$ are set to zero with a variance equal to 1/$N$.
$N, \beta_x$, and $\beta_y$ are set to $100, 2.0$, and $2.0$, respectively, while the time scales of $x$ and $y$, denoted as $\tau_{x}$ and $\tau_{y}$, are set to 1 and 33, respectively.
$I_{i}$ is the $i$ th element of an input pattern.

\subsection{Learning process}
In our model, only $J^{X}$ is plastic and changes according to 
\begin{equation*}
	\dot{J_{ij}^{X}} = \epsilon(\xi_i - x_i)(x_j - u_i J_{ij}^X)/N,
\end{equation*}
where $\epsilon$ is the learning speed, and it is set to $0.03$.
$\xi_i$ is the $i$-th element of a target $\boldsymbol{\xi}$, which is an $N$-dimensional pattern.
In previous studies\cite{Kurikawa2013,Kurikawa2016,Kurikawa2020}, we demonstrated that a single population with a single timescale learns mappings between constant input and target patterns using this learning rule.
In the current two-population model, there are two inputs for the fast sub-network---one is from an external input and the other is from the slow sub-network that stores previous information.
Thus, the synaptic dynamics can modify the connection to generate a target pattern depending not only on the currently applied input but also on the preceding input, as shown in \cite{Kurikawa2020a}.
We, in the present study, demonstrate this model performs a context-dependent task.

\begin{figure}
\includegraphics[width=\textwidth]{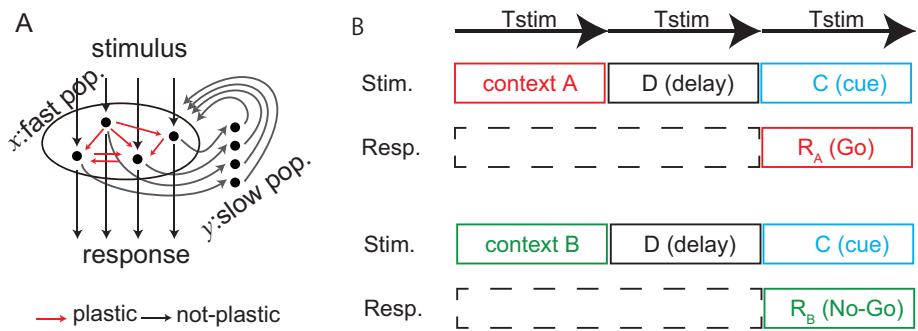}
\caption{A: Schematic of our model. B: Context-dependent task. $T_{stim}$ denotes the stimulus duration.}
\label{fig:fig1}
\end{figure}

\section{Results}
\subsection{Context-dependent task}
First, we consider learning a simple context-dependent task that is composed of two context signals, a delayed signal and a cue signal (Fig. \ref{fig:fig1}B). 
In this task, one of the context signals is applied to fast neurons, followed by the delayed signal.
Finally, the cue signal is applied.
On applying the cue signal, the network is required to generate ‘‘Go'' pattern or ‘‘No-Go'' pattern depending on the context signal.
When the applied context signal is $A$, the network should generate the Go pattern, while it should generate the No-Go pattern when the context signal is $B$.
Here, we denote Go and No-Go patterns as $R_A$ and $R_B$, respectively.
Thus, the network must maintain the context signal to generate an adequate response pattern.

The signal and response patterns are random $N$-bit binary patterns, each element of which corresponds to a neuron of $x$, 
with probabilities $P(X_{i}=\pm 1)= 1/2$, where $X_{i}$ is the activity state of the $i$-th element of patterns $X=A,B,C, R_A,$ and $R_B$.
All elements of the delayed signal $D$ are $-1$.
We apply the context, the delayed, and the cue signal sequentially with duration time $T_{stim}=60$ as an external input $\eta$ in Eq. \ref{eq:neuro-dyn}, as illustrated in Fig. \ref{fig:fig1}B.
If the fast dynamics reach $R_A$ (or $R_B$), namely $\sum {R_A}_i x_i /N>0.95$, a trial is completed, and the next trial starts.
Now, the target patterns are defined only when the cue signal is applied.
Therefore, the learning process runs only during the cue signal application; otherwise, only the neural dynamics run.

\begin{figure}
\includegraphics[width=\textwidth]{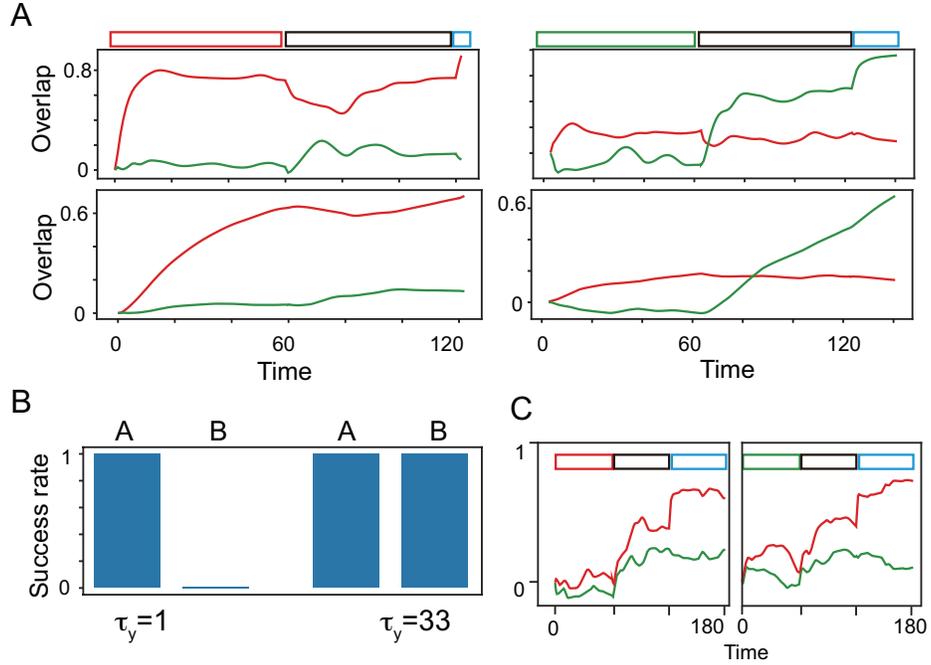}
\caption{A: Neural dynamics after being trained in the context-dependent task for the context A and B in the left and right panels, respectively. The neural dynamics of $x$ and $y$ are plotted by computing their overlaps with $R_A$ (red line) and $R_B$ (green line), respectively. The top panels represent the neural dynamics of $x$, whereas the bottom panels represent the neural dynamics of $y$.
The colored bars above the top panels represent context signals $A$ and $B$, the delay signal, and the cue signal in red, green, black, and cyan, respectively.
B: Success rate of the context-dependent task for $\tau_y = 33$ and $1$. For each condition, we plot the success rate of the response to contexts A and B separately.
C: Neural dynamics for $\tau_y=1$ for the context A and B in the left and right panels, respectively.
The neural trajectories are plotted in the same manner as in panel $A$.}
\label{fig:fig2}
\end{figure}

We show successful trials in the context-dependent task in Figure \ref{fig:fig2}A.
In the presence of the context signal $A$, fast neural dynamics already show a high overlap with $R_A$, while they show quite different patterns in the presence of $B$. 
The slow dynamics follow the fast dynamics in both cases.
The fast dynamics change slightly upon the delayed signal after the context signal $A$, while they change drastically after $B$.
Although the network receives the same delayed signals, quite different neural dynamics emerge owing to the slow dynamics that reflect the preceding context signals.
Finally, the fast dynamics converge to the correct target $R_A$ or $R_B$ depending on the preceding context signals $A$ and $B$, respectively.
We measured the success rate across 20 trials for each context signal, resulting in a 95\% success rate ( Fig. \ref{fig:fig2}B). 

To examine the role of the slow dynamics in this task, we analyzed the neural dynamics with $\tau_y=1$, i.e., the dynamics of the two populations change with the same timescale.
In this case, the neural dynamics $y$ do not store the history of the fast dynamics $x$ over a long time.
We plot two trajectories for the context A and B in Fig. \ref{fig:fig2}C.
In the present of the context signals, two trajectories show different behaviors depending on the identity of the context signals.
After the presentation of the context signal, the difference between the two trajectories rapidly decreased during the delayed epoch and the cue one.
Thus, the network generates the same neural patterns (in this case, $R_A$) in both contexts $A$ and $B$ in the presence of the cue signal. Altogether, these results show that the the population with the slow timescale enables successful performing the context-dependent task.

\subsection{Delayed match to sample task}
\begin{figure}
\includegraphics[width=\textwidth]{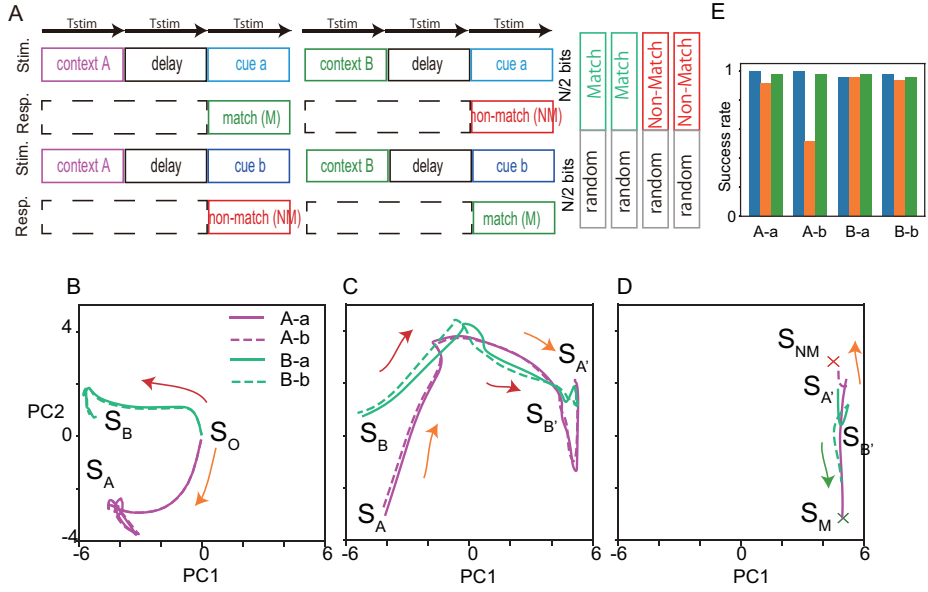}
\caption{A: Schematic of the delayed match to sample task. B--D: Fast dynamics are plotted on two principal component spaces (B, C, and D: the dynamics in the presence of the context signal, the delayed one, and the cue one, respectively). The four trajectories represent fast dynamics under different conditions. E: Success rate against noise and perturbation for each condition. The performances for normal, perturbation in initial states, and modified $T_{stim}$ situations are shown in blue, orange, and green, respectively.} \label{complex_task_1}
\label{fig:fig3}
\end{figure}

In the task analyzed above, the network is required to store only the context information irrespective of the cue signals; this is because the correct response is uniquely determined by the context signal.
However, the required responses are commonly dependent on cue signals in addition to context signals, as analyzed in \cite{Mante2013,Stokes2013}.
Subsequently, we examined whether our model performs such a complex context-dependent task (denoted as a delayed match to the sample task), as shown in Fig. \ref{fig:fig3}A.
This task is based on procedure similar to the task analyzed above.
There are two context signals, $A$ and $B$, and two cues, $a$ and $b$, and the same-name patterns (i.e., $A$ and $a$, or $B$ and $b$) belong to the same category.
A network is required to ‘‘Go'' (denoted as matched pattern $M$) when the context and cue signal in the same category are given, while it is required to ‘‘No-Go'' (denoted as non-matched pattern $NM$) when these cues are in a different category.
Unlike the previous task, the network should generate two different patterns depending on the cue signals, in addition to the context signals.

The context and cue signals ($A, B, a$, and $b$) and response patterns ($M$ and $NM$) are the random $N$-bit patterns generated in the same manner as in the previous task.
There are two $M$ patterns for $A-a$ and $B-b$ as well as two $NM$ patterns for $A-b$ and $B-b$.
When the two $M$ ($NM$) patterns are the same, the learning performance is poor.
Thus, we add random perturbation parts to $M$ and $NM$, as shown in Fig. \ref{fig:fig3}A.
The delay signal and learning procedure are the same as that in the previous task.

First, we present four trajectories corresponding to the four conditions (two contexts by two cues) in Figs. \ref{fig:fig3}B-D.
Here, the fast neural dynamics are projected onto the 2D principal component (PC) space after averaging over 20 trajectories for each condition.
The projected trajectories for four different conditions are shown by solid and dotted lines in different colors.

In the presence of context signals, neural trajectories are separated into two groups according to the context signals and converge into neural states $S_A$ and $S_B$.
Following the context signals, the delay signals are applied.
Two groups of trajectories evolve from $S_A$ and $S_B$ and reach $S_A'$ and $S_B'$ with a decrease in the distance between the two groups.
Although the difference decreases, that between neural states in the slow dynamics remains high at the end of the delay epoch, and the information of the preceding context signal is retained in the slow dynamics.
Finally, applying the cue to the network separates the four trajectories into the match ($S_M$) and non-match ($S_{NM}$) states depending on the contexts and cues.
The trajectories diverge into different states in the presence of the same cue owing to the slow dynamics that reflect the context signals. 
In this manner, the network performs the delayed match to sample task.
To precisely evaluate the performance of this model, we measured the success rate for 20 trials for each condition, resulting in more than $90$ \% trials being successful for all conditions.

In a noisy system such as a neural system, information processing should be robust against noise and perturbations.
We explore the robustness of our model against perturbations in the initial states and the change in stimulus duration $T_{stim}$.
First, for the modification of $T_{stim}$, we changed $T_{stim}$ for each epoch (the context signal, the delay signal, and the cue signal) from $T_{stim}=60$ to $T_{stim}=66$ in the test process after training the network with $T_{stim}=60$.
We measured the success rate by 20 trials for modified $T_{stim}$ and observe that such perturbations do not reduce the success rate compared to the normal situation, as shown in Fig. \ref{fig:fig3}E.

\begin{figure}
\includegraphics[width=\textwidth]{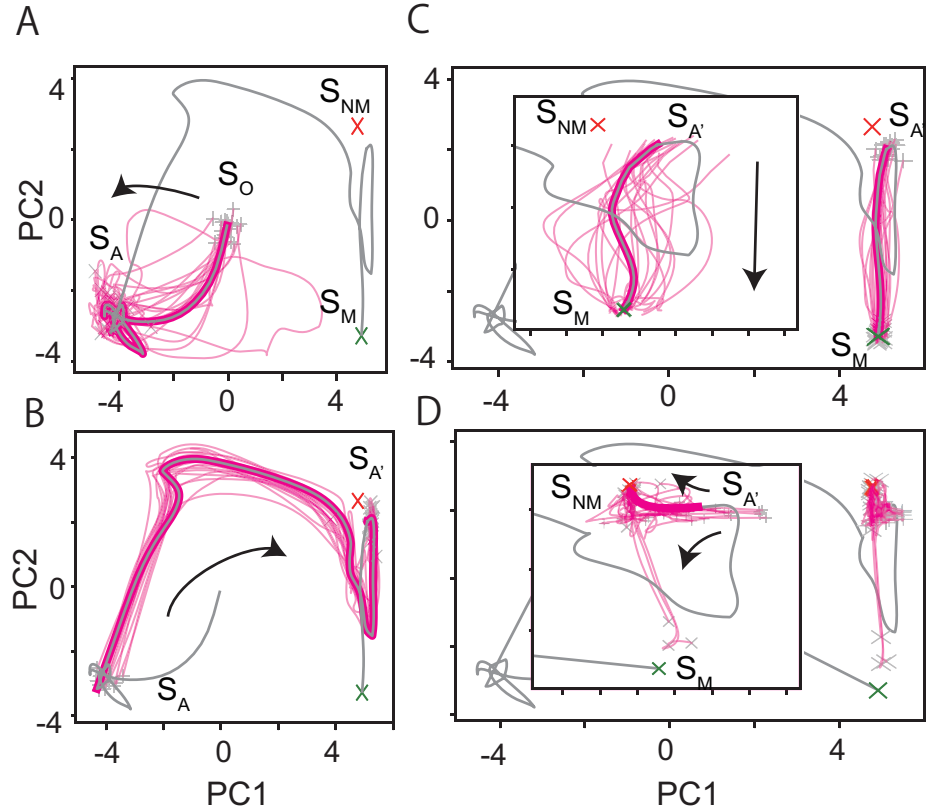}
\caption{A--C: Neural trajectories with perturbed initial states projected on 2D PC space for context A--cue a condition.
Trajectories during the context signal, the delayed one, and the cue one are plotted on A, B, and C, respectively. D: Neural trajectories during the cue signal for the context A--cue b condition.} 
\label{fig:fig4}
\end{figure}

Next, we analyze the robustness against the perturbations of the initial states, wherein the initial states of the fast and slow dynamics, $\boldsymbol{x}$ and $\boldsymbol{y}$, in the test process are selected randomly from a larger distribution around the origin in the neural state space than those in the learning process.
We set the initial states of $\boldsymbol{x}$ and $\boldsymbol{y}$ in the test process to random $N$-dimensional states uniformly sampled from a closed interval $[-0.2,0.2]^N$, 
whereas in the learning process, the initial states are limited to a smaller closed interval $[-0.01,0.01]^N$.
Therefore, almost all initial states in the test process are novel for the network.
By measuring the success rate against 20 trials for the perturbation case, we observed that $90$ \% of all trials were successful for three conditions, while more than $50$ \% of all trials were successful for one condition (context A -- cue b).
The network successfully performs the delayed match to sample task with highly volatile initial states, except for A – b condition.
In total, these results indicate that our learning process generates a network model robust against perturbations of the initial states and signal duration.

Why is such robust information processing possible?
To answer this question, we analyzed the stability of neural trajectories in each epoch (i.e., in the presence of context, delayed, and cue signals) in the test process.
First, to explore the robustness during the signal epoch, the initial states are randomly generated around $S_O$ in the same manner as the perturbation case, and the fast and slow dynamics run according to Eqs. \ref{eq:neuro-dyn} and \ref{eq:neuro-dyn-slow}.
Figure \ref{fig:fig4}A demonstrates these trajectories in the presence of the context $A$.
Neural trajectories from the perturbed initial states converge to and remain around $S_A$ for some time.
$S_A$ is stable and attracts trajectories starting from a broad area of the neural state space in the context $A$ epoch.
The same analysis in the delayed and cue epochs revealed that $S_B, S_A', S_B', S_M$, and $S_{NM}$ are stable states in the corresponding epochs, as shown in Figs. \ref{fig:fig4}B and C.
After changing the signal, each state is destabilized.
$S_A'$ and $S_B'$ are stable for the same delay signals depending on the preceding contexts.
$S_M$ and $S_{NM}$ are stable dependent on the context and the cue.
The slow dynamics store the context information and stabilize the different states through interaction with the cue signal.
These results demonstrate that successive stabilization and destabilization of the epoch-specific states underlie the stable performance of the delayed match to the sample task.

In contrast, $S_{MN}$ in the presence of $b$ (i.e., the correct target state for A – b condition) is less stable than the other task-related states, resulting in some trajectories not reaching $S_{NM}$, but $S_M$, as shown in Fig. \ref{fig:fig4}D.
Thus, the task often fails. 
These results indicate that our model robustly performs a delayed match to sample task by sequential stabilization of the task-related states.

\section{Discussion and Summary}
We have proposed and demonstrated that sequential stabilization of task-related states leads to the robust performance of the context-dependent task.
During each epoch (i.e., in the presence of the context signal, the delayed one, and the cue one), the epoch-specific states are stabilized and destabilized in the fast dynamics through the regulation of the slow dynamics.
The slow dynamics maintain the previous stimulus and enable the neural state in the fast dynamics to transit from one state to another in context-dependent (i.e., non-Markov) manner.

Such transitions of neural states are distinguished from the typical models of the transitions.
Although asymmetric Hebbian connection model\cite{Amari1972,Kleinfeld1986,Sompolinsky1986,Seliger2003,Recanatesi2015,Russo2012,Gros2007} generates transitions between metastable states, the current pattern is determined only by the immediately preceding pattern and, consequently, non-Markov sequences are not allowed. 
Sophisticated training methods of RNNs\cite{Mante2013,Chaisangmongkon2017,Sussillo2009,Laje2013} generate complex neural trajectories, but how such trajectories are formed is unclear and the stability of these trajectories is poor\cite{Laje2013}.

Recent theoretical studies revealed relevance of the stable states in neural processing\cite{Sussillo2015,Sussillo2013}.
In these studies, several stable states exist in the neural state space and the adequate input drives the neural state from one stable state to another for performing the cognitive tasks, e.g., flip-flop function and generation of muscle activity.
Thus, initial states before being applied the input are crucial for performing a task.
In contrast, in our model, the input stabilizes a single (or a few) state(s) in the fast dynamics depending on the previous inputs stored in the slow dynamics.
Hence, neural states beginning from a broader area converge the adequate state.
Our proposed model sheds light on neural processing by the hierarchical timescale structure, which are commonly observed in the neural system\cite{Murray2014,Runyan2017}.

\section*{Acknowledgments}
We thank Kunihiko Kaneko for fruitful discussion for our manuscript.
This work was partly support by JSPS KAKENHI (no.20H00123).

%
%
%
%
\bibliographystyle{splncs04}
\bibliography{13th}

\end{document}